\documentclass[aps,amsmath,amssymb]{revtex4}

\usepackage[dvips]{graphicx}

\usepackage{dcolumn}
\usepackage{bm}
 
\begin{document}

\title{Tidal gravitational effects in a satellite.}
\author{Ph. Tourrenc}
\author{M-C. Angonin}
\author{X. Ovido}
\affiliation{
Universit\'{e} P. et M. Curie\\
ERGA, case 142\\
4, place Jussieu\\
F-75252 Paris CEDEX 05, France}

\date{\today}

\begin{abstract}
Atomic wave interferometers are tied to a telescope pointing towards a
faraway star in a nearly free falling satellite. Such a device is sensitive
to the acceleration and the rotation relatively to the local inertial frame
and to the tidal gravitational effects too.

We calculate the rotation of the telescope due to the aberration and the
deflection of the light in the gravitational field of a central mass (the
Earth and Jupiter). Within the framework of a general parametrized
description of the problem, we discuss the contributions which must be taken
into account in order to observe the Lense-Thirring effect.

Using a geometrical model, we consider some perturbations to the idealized
device and we calculate the corresponding effect on the periodic components
of the signal.

Some improvements in the knowledge of the gravitational field are still
necessary as well as an increase of the experimental capabilities~; however
our conclusions support a reasonable optimism for the future.

Finally we put forward the necessity of a more complete, realistic and
powerful model in order to obtain a definitive conclusion on the feasibility
of the experiment as far as the observation of the Lense-Thirring effect is
involved.
\end{abstract}

\maketitle

\section{Introduction}

Clocks, accelerometers and gyroscopes based on cold atom
interferometry are already among the best which have been constructed until
now and further improvements are still expected. With the increase of the
experimental capabilities it becomes necessary to consider more and more
small effects in order to account for the signal, therefore (relativistic)
gravitation has to be considered in any highly sensitive experiments, no
matter what they are designed for.

The performances of laser cooled atomic devices is limited on Earth by
gravity. Further improvements demand now that new experiments take place in
free falling (or nearly free falling) satellites. A laser cooled atomic
clock, named PHARAO, will be a part of ACES (Atomic Clock Ensemble in
Space), an ESA mission on the ISS. Various other experimental possibilities
involving \textquotedblright Hyper-precision cold atom interferometry in
space\textquotedblright\ are presently considered. They might result in a
project (called \textquotedblright Hyper\textquotedblright ) in the
future \cite{Raselhyper}.

The aim of the present paper is to hold the bookkeeping of the various
gravito-inertial effects in a nearly free falling satellite. For this
purpose we consider the most ambitious goal which has been considered for
Hyper \textit{i.e.} the measurement of the Lense-Thirring effect.

The Lense-Thirring effect is a local rotation of a gyroscope relatively to a
telescope pointing towards a far away star. It is a relativistic consequence
of the diurnal rotation of the Earth which "drags the inertial frames" in
its neighborhood.

The angular velocity of the telescope relative to the gyroscopes depends
on the position. Therefore, in a satellite, it is a function of the time. In
Hyper, the angular velocity is measured by atomic-wave-gyroscopes and its
time dependence is analyzed \footnote{The Lense-Thirring
effect results also in a secular precession which is not considered here but
in Gravitational Probe B~: a NASA\ experiment which is planed to be launched
on the 6th of December 2003.}. The consequence is that the device is
sensitive to the variation of the gravitation in the satellite and no to the
gravitation itself. We do not believe that it is easy to achieve the
required stabilization of the gravitational field due to the local masses
but it is not impossible in principle. For this reason we will study only
the tidal field of far away masses whose effect cannot be removed at all.

The parameter which plays a role in the calculation of the Lense-Thirring
effect is the angular momentum of the central mass. It is much bigger for
Jupiter than the Earth. Therefore we will discuss both cases, without any
consideration on the cost of the corresponding missions.

In the sequel the greek indices run from $0$ to $3$ and the Latin indices
from $\ 1$ to $3.$ We use the summation rule of repeated indices (one up and
one down).

The Minkowski tensor is $\eta _{\alpha \beta }=diag\left[ 1,-1,-1,-1\right] $%
; its inverse is $\eta ^{\alpha \beta }.$

We use geometrical units where $c=G=1.$

\section{The local experiment in a satellite}

\begin{center}
\begin{figure}[ht]
\centering\includegraphics[width = .75\linewidth]{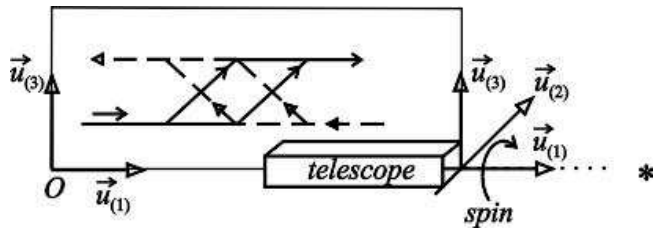}
\caption{\label{asu} The experimental setup.}
\end{figure}
\end{center}

In the satellite, the experimental set-up consists in a telescope pointing
towards a far away star in the $\overrightarrow{u}_{\left( 1\right) }$
direction and two orthogonal atomic Sagnac units in the planes $\left[
\overrightarrow{u}_{\left( 3\right) },\overrightarrow{u}_{\left( 1\right) }%
\right] $ and $\left[ \overrightarrow{u}_{\left( 2\right) },\overrightarrow{u%
}_{\left( 1\right) }\right] $ of fig.\,\ref{asu}.

\subsection{The atomic Sagnac unit}

\begin{center}
\begin{figure}[ht]
\centering\includegraphics[width = .75\linewidth]{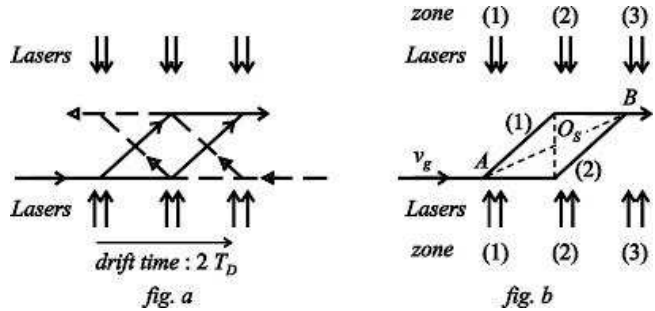}
\caption{\label{figasu} An atomic Sagnac unit (ASU).}
\end{figure}
\end{center}

An atomic Sagnac unit (ASU) is made of two counter-propagating atom
interferometers which discriminate between rotation and acceleration (see
figure \ref{figasu}-$a$).

Each interferometer is a so-called Ramsey-Bord\'{e} interferometer with a
Mach-Zehnder geometry ( figure \ref{figasu}-$b$). The atomic beam from a
magneto-optical trap interacts three times with a laser field. In the first
interaction zone the atomic beam is split coherently, by a Raman effect,
into two beams which are redirected and recombined in the second and the
third interaction zone.

The mass of the atom depends on its internal state, therefore it is not a
constant along the different paths. However, the change of the mass is very
small; it leads to negligible corrections on the main effects which is
already very small. In the case of the cesium, the mass is $m=133\times
1.66\times 10^{-27}=2.2\times 10^{-25}\text{kg}$ and the wave length of the
lasers is $\lambda =850\text{nm}.$ The momentum transferred to the atom
during the interaction is $\dfrac{4\pi \hbar }{\lambda }.$ The recoil of the
atom results in a Sagnac loop which permits to measure the angular velocity
of the set-up relatively to a local inertial frame. The device is also
sensitive to the accelerations.

In an ideal set-up the two interferometers are identical coplanar
parallelograms with their center $O_{S}$ and $O_{S}^{\prime }$ at the same
point but many perturbations have to be considered. The geometry of the
device is actually determined by the interaction between the initial atomic
beam and the lasers~; Therefore a full treatment of the atom-laser interaction
in a gravitational field is obviously necessary to study the response of the
Atomic Sagnac Unit (ASU). 
However the geometrical model is useful to give a physical intuition of
the phenomena. In this context we assume that the two interferometers remain
idealized identical parallelograms but that $O_{S}$ and $O_{S}^{\prime }$
are no longer at the same point~: This is the only perturbation that we
consider here. It is sufficient to take the flavor of the gravitational
perturbations which have to be taken into account and, more generally, of
the difficulty inherent to such an experiment.

\subsection{The phase difference}

\label{phasediff}

Let us assume that the fundamental element is known in some coordinates
comoving with the experimental set-up~:%
\begin{eqnarray}
ds^{2} & = & \left( 1+K_{\left( 0\right) \left( 0\right) }\right)
dT^{2}+2K_{\left( 0\right) \left( k\right) }dTdX^{\left( k\right) }\\
&&+\left(
\eta _{\left( k\right) \left( j\right) }+K_{\left( k\right) \left( j\right)
}\right) dX^{\left( k\right) }dX^{\left( j\right) } \nonumber
\label{decadix}
\end{eqnarray}

In order to calculate up to first order the gravitational perturbation of
the phase due to $K_{\left( \alpha \right) \left( \beta \right) },$ we use a
method which we summarize now \cite{LetT}.

First we calculate the quantity $\Psi $~:%
\begin{equation}
\Psi =K_{\left( 0\right) \left( 0\right) }+2K_{\left( 0\right) \left(
k\right) }v_{g}^{\left( k\right) }+K_{\left( k\right) \left( j\right)
}v_{g}^{\left( k\right) }v_{g}^{\left( j\right) }  \label{psipsi}
\end{equation}%
where $v_{g}^{\left( k\right) }$ is the velocity of the atoms (\textit{i.e.}
the unperturbed group velocity).

The quantity $\Psi $ is a function of the time and the position of the atom.

Then we consider an atom which arrives at time $t$ at point $B$ of figure %
\ref{figasu}-$b~.$ Now the position is a function of the time $t^{\prime }$
only because $t$ is considered as a given quantity. The function $\Psi $ is
a function of the time, $t^{\prime },$ only. The phase difference is

\begin{equation}
\delta \varphi =\dfrac{\omega }{2}\int_{t-2T_{D}/(2)}^{t}\Psi \,(t^{\prime
})\;dt^{\prime }-\dfrac{\omega }{2}\int_{t-2T_{D}/\left( 1\right) }^{t}\Psi
\,(t^{\prime })\;dt^{\prime }  \label{deltafi}
\end{equation}%
The integrals are performed along path (2) and (1) of the
interferometer (figure \ref{figasu}-$b$)%
. The \textquotedblright angular frequency\textquotedblright\ $\omega $ is
defined as $\dfrac{m\,c^{2}}{\hbar }.$

\subsection{The local metric}

In order to calculate $\delta \varphi ,$ we must know the local metric $%
G_{\left( \alpha \right) \left( \beta \right) }=\eta _{\left( \alpha \right)
\left( \beta \right) }+K_{\left( \alpha \right) \left( \beta \right) }.$

We choose an origin, $O,$ in the satellite and at point $O$ a tetrad $%
\left\{ u_{\left( 0\right) }^{\alpha },u_{\left( 1\right) }^{\alpha
},u_{\left( 2\right) }^{\alpha },u_{\left( 3\right) }^{\alpha }\right\} $
where $u_{\left( 0\right) }^{\alpha }$ is the 4-velocity of point $O$ and
where the three vectors $\left\{ u_{\left( k\right) }^{\alpha }\right\}
=\left\{ 0,\overrightarrow{u}_{\left( k\right) }\right\} $ are represented
on the figure \ref{asu}. The vectors of the tetrad are orthogonal~: $%
u_{\left( \mu \right) }^{\alpha }u_{\alpha \left( \sigma \right) }=\eta
_{\left( \mu \right) \left( \sigma \right) }.$ The coordinate indices, $\alpha
,\beta ,\sigma ,$ etc, are lowered or raised by the means of the metric
tensor, $g_{\alpha \beta }$ or $g^{\alpha \beta }~;$ the Minkowski indices
are raised or lowered with the Minkowski metric or its inverse, $\eta
_{\alpha \beta }$ or $\eta ^{\alpha \beta }$. An Einstein indice, $\alpha $,
can be changed into a Minkowski indice $\left( \rho \right) $, by the means
of the tetrad and vice versa~: $u_{\left( \rho \right) }^{\alpha }\left( \
\right) _{\alpha }=\left( \ \right) _{\left( \rho \right) }$ and $u_{\left(
\rho \right) }^{\alpha }\left( \ \right) ^{\left( \rho \right) }=\left( \
\right) ^{\alpha }.$

The tetrad is the natural basis at point $O$ of comoving coordinates $%
X^{\left( \alpha \right) }.$ The proper time at the origin is $X^{\left(
0\right) }.$ The space coordinates are the $X^{\left( k\right) }.$

Once the origin and the tetrad are chosen, the metric at point $M$ and time $%
t$ is expanded relatively to the space coordinates of $M$ \cite{LiNi}.

\begin{eqnarray}
ds^{2} &=&G_{\left( \alpha \right) \left( \beta \right) }\,dX^{\left( \alpha
\right) }dX^{\left( \beta \right) }\text{\ \ with}  \label{local} \\
&&\;  \notag \\
G_{\left( 0\right) \left( 0\right) } &=&1+2\,\vec{a}\cdot \vec{X}+\left(
\vec{a}\cdot \vec{X}\right) ^{2}-\left( \vec{\Omega}\times \vec{X}\right)
^{2}-R_{\left( 0\right) \left( k\right) \left( 0\right) \left( j\right)
}\,X^{\left( k\right) }X^{\left( j\right) }  \notag \\
&&-\dfrac{1}{3}R_{\left( 0\right) \left( k\right) \left( 0\right) \left(
j\right) ,\left( \ell \right) }X^{\left( k\right) }X^{\left( j\right)
}X^{\left( \ell \right) }+... \\
G_{\left( 0\right) \left( m\right) } &=&\Omega _{\left( m\right) \left(
k\right) }\,X^{\left( k\right) }-\dfrac{2}{3}R_{\left( 0\right) \left(
k\right) \left( m\right) \left( j\right) }\,X^{\left( k\right) }X^{\left(
j\right) } \notag \\
&&-\dfrac{1}{4}R_{\left( 0\right) \left( k\right) \left( m\right)
\left( j\right) ,\left( \ell \right) }X^{\left( k\right) }X^{\left( j\right)
}X^{\left( \ell \right) }+..  \notag \\
G_{\left( n\right) \left( m\right) } &=&\eta _{\left( n\right) \left(
m\right) }-\dfrac{1}{3}R_{\left( n\right) \left( k\right) \left( m\right)
\left( j\right) }X^{\left( k\right) }X^{\left( j\right) } \notag\\
&&-\dfrac{1}{6}%
R_{\left( n\right) \left( k\right) \left( m\right) \left( j\right) ,\left(
\ell \right) }X^{\left( k\right) }X^{\left( j\right) }X^{\left( \ell \right)
}+...  \notag
\end{eqnarray}%
where we have used vector notations \textit{i.e.} $\overrightarrow{a}$ for $%
\left\{ a^{\left( \ell \right) }\right\} ,\ \overrightarrow{a}\cdot
\overrightarrow{X}$ for $\sum a^{\left( \ell \right) }\,X^{\left( \ell
\right) },$ etc. Every quantity, except the space coordinates $X^{\left(
\ell \right) },$ are calculated at point $O.$ Thus they are functions of the
time $T=X^{\left( 0\right) }.$

$R_{\left( \alpha \right) \left( \beta \right) \left( \sigma \right) \left(
\mu \right) }$ is the Riemann tensor obtained from $R_{\alpha \beta \sigma
\mu }$ at point $O:$%
\begin{equation}
R_{\alpha \beta \sigma \mu }=\Gamma _{\alpha -\beta \mu ,\sigma }-\Gamma
_{\alpha -\beta \sigma ,\mu }+\Gamma _{\beta \sigma }^{\varepsilon }\Gamma
_{\varepsilon -\alpha \mu }-\Gamma _{\beta \mu }^{\varepsilon }\Gamma
_{\varepsilon -\alpha \sigma }  \label{riem}
\end{equation}%
where $\Gamma _{\alpha -\beta \mu ,\sigma }$ is the Christoffel symbol.

$\Omega _{\left( j\right) \left( k\right) }$ is the antisymmetric quantity
\begin{eqnarray}
\Omega _{\left( j\right) \left( k\right) } &=&\dfrac{1}{2}\left( g_{\left(
0\right) \left( j\right) ,\left( k\right) }-g_{\left( 0\right) \left(
k\right) ,\left( j\right) }\right) _{O}  \label{omega} \\
&&\;\;\;\;\;\;\;\;\;+\dfrac{1}{2}\left( \left( u_{\left( j\right) }^{\beta
}\,\dfrac{du_{\left( k\right) }^{\alpha }}{ds}-\dfrac{du_{\left( j\right)
}^{\beta }}{ds}\,u_{\left( k\right) }^{\alpha }\right) g_{\alpha \beta
}\right) _{O}  \notag
\end{eqnarray}

Due to the antisymmetry of $\Omega _{\left( m\right) \left( k\right) },$ the
quantity $\Omega _{\left( m\right) \left( k\right) }\,X^{\left( k\right)
}dX^{\left( m\right) }$ which is present in the expression of $ds^{2}$ can
be written as $\Omega _{\left( m\right) \left( k\right) }\,X^{\left(
k\right) }dX^{\left( m\right) }=\left( \overrightarrow{\Omega }_{0}\wedge
\overrightarrow{X}\right) \cdot d\overrightarrow{X}$. The space vector $%
\overrightarrow{\Omega }_{0}$ is the physical angular velocity. It is
measured by gyroscopes tied to the three space orthonormal vectors $%
u_{\left( k\right) }^{\alpha }$~:

The vector $\overrightarrow{a}$ is the physical acceleration which can be
measured by an accelerometer comoving with $O.$ It is the spatial projection
at point $O$ of the 4-acceleration of point $O.$

At point $O$ ($i.e.$ $\overrightarrow{X}=\overrightarrow{0})$ the time $T$
is the proper time delivered by an ideal clock comoving with $O.$

\section{From the geocentric coordinates to the comoving coordinates}

\subsection{The geocentric coordinates}

We define the time coordinate $x^{0}=ct$ and the space coordinates $x^{k}$.
We use the notations $\vec{r}=\left\{ x^{k}\right\} =\left\{ x,y,z\right\} $
and we define the spherical coordinates $\left\{ r,\theta ,\varphi \right\}
, $ \textit{i.e. }$x=r\sin \theta \,\cos \varphi \;,\;\;y=r\sin \theta
\,\sin \varphi \;,\;\;z=r\cos \theta .$

We consider a satellite and a point $O$ which is the origin of the local
coordinates in the satellite. We assume that the position of $O$ is given by
its three space coordinates, $\vec{r}=\left\{ x,y,z\right\} =\left\{
x^{k}\right\} ,$ considered as three known functions of the coordinate time,
$t$. Then we define the velocity of point $O$ as $\overrightarrow{v}=\dfrac{d%
\vec{r}}{dt}.$

The proper time at point $O$ is $s=T=X^{\left( 0\right) }.$ The motion of $O$
can be described as well by the four functions $x^{\alpha }=x^{\alpha
}\left( s\right) .$ The four-velocity is defined as $u^{\alpha }=\dfrac{%
dx^{\alpha }}{ds}.$

In the sequel we consider the \textsc{P}arametrized \textsc{P}ost \textsc{N}%
ewtonian theories \cite{will}. The relevant
\textsc{PPN} parameters which appear below are $\gamma $ and $\alpha _{1}.$
The parameter $\gamma $ is the usual parameter connected to the deflection
of a light ray by a central mass. The parameter $\alpha _{1}$ couples the
metric to the speed, $-\vec{w}$, of the preferred frame (if any) relatively
to the geocentric frame. In general relativity, $\alpha _{1}=0$ and $\gamma
=1.$

Let us define now several quantities which will be used in the sequel~:

$\bullet \qquad 2M_{\circledS }$ is the Schwarzschild's radius of the
central body (\textit{i.e.} the Earth or Jupiter). As we use geometrical
units, $M_{\circledS }$ is also its \textquotedblright
mass\textquotedblright .

$\bullet \qquad \vec{J}_{\circledS }$ is the angular momentum of the central
body in geometrical units$.$ The relevant quantity which appears below, is $%
\vec{J}=\dfrac{1+\gamma +\alpha _{1}/4}{2}\vec{J}_{\circledS }$. We define $%
J=\left\Vert \vec{J}\right\Vert \simeq \left\Vert \vec{J}_{\circledS
}\right\Vert =J_{\circledS }$

$\bullet \qquad \vec{g}=-2\;\dfrac{\vec{J}\wedge \vec{r}}{r^{3}}+\dfrac{1}{2}%
\alpha _{1}U\;\vec{w}$ is the definition of $\vec{g}$, where $\vec{w}$
is the velocity of the observer, relative to the preferred frame (if any).

$\bullet \qquad U$ is the Newtonian potential
\begin{equation}
U=\dfrac{M_{\circledS }}{r}\left( 1-J_{2}\left( \dfrac{R_{\circledS }}{r}%
\right) ^{2}P_{2}+\Delta \right) +U_{\ast }  \label{4pole}
\end{equation}%
where $R_{\circledS }$ is the radius of the central body and $U_{\ast }$ the
potential due to its satellites, the Sun and the planets\footnote{%
An arbitrary constant can always be added to $U_{\ast }.$ It is chosen in
such a way that zero is the mean value of $U_{\ast }$ at point $O$ in the
satellite.}. In spherical coordinates the Legendre polynomial $P_{2}$ reads $%
P_{2}=\dfrac{1}{2}\left( 3\cos ^{2}\theta -1\right) .$ The quadrupole
coefficient is $J_{2}$ and $\Delta $ represents the higher harmonics. It
depends on the angle $\varphi $ and on the time $t$ because of the rotation
of the central body.

In the non rotating geocentric coordinates the significant fundamental
element is
\begin{equation}
ds^{2}=\left( 1-2U\right) dt^{2}+2g_{0k}\,dx^{k}dt-\left( 1+2\gamma U\right)
\delta _{jk}\,dx^{j}dx^{k}  \label{ds2}
\end{equation}%
where $\left( \vec{g}\right) _{k}=-\left( \vec{g}\right) ^{k}=g_{0k}.$ In
eq.\,(\ref{ds2}), we have dropped post Newtonian corrections which
are too small to be considered here.

\subsection{Orders of magnitude}

Table 1 below gives the order of magnitude of the various parameters which
have been introduced previously.

\begin{center}
\begin{tabular}{c}
\begin{tabular}{|l|l|l|l|l|l|}
\hline
& $M_{\circledS }$ & $J_{\circledS }$ & $R_{\circledS }$ & $J_{2}$ & $\Delta
$ \\ \hline
Earth & $4.4\text{mm}$ & $145\text{cm}^{2}$ & $6400\text{km}$ & $\sim
10^{-3} $ & $\sim 10^{-6}$ \\ \hline
Jupiter & $1.4\text{m}$ & $1700\text{m}^{2}$ & $71300\text{km}$ & $\sim
10^{-2}$ & $\lesssim 10^{-3}$ \\ \hline
\end{tabular}
\\
Table 1.%
\end{tabular}
\end{center}

In order to describe the physical situation we introduce four\ parameters~:~$%
\xi ,$ $\varepsilon ,$ $\eta $ and $\mu .$

First we define the order of magnitude $O_{1}=\sqrt{\dfrac{M_{\circledS }}{%
R_{\circledS }}}.$ The quantity $\left( O_{1}\right) ^{n}$ is denoted
by $O_{n}$.

Then we consider a nearly free falling satellite on a nearly circular orbit
of radius $r\sim R_{\circledS }/\xi $. This expression gives the definition
of $\xi .$ The velocity of the satellite is of order $v=\xi ^{1/2}O_{1}$
\footnote{Notice that the quantity $\xi ^{1/2}O_{1}$ is what is
called $O_{1}$ in Will's book quoted above}.

Now we define $d=R_{\circledS }O_{1}$ and $\varepsilon $ such as $%
X=\varepsilon d$ where $X$ is the size of the laboratory.

We define $\eta .$ The velocity of the atoms is $v_{g}=\eta O_{1}$.

Finally we assume that the various quantities such as the position of $O$ or
the geometry of the experimental set-up is known with a relative accuracy of
order of $\mu .$

\begin{center}
\begin{tabular}{c}
\begin{tabular}{||c||c||}
\hline\hline
$O_{1}=\sqrt{\dfrac{M_{\circledS }}{R_{\circledS }}}$ & relative accuracy~:~$%
\mu $ \\ \hline\hline
Orbital parameters & set-up parameters \\ \hline
\begin{tabular}{ll}
radius & $r=\dfrac{R_{\circledS }}{\xi }$ \\
velocity & $v=O_{1}\xi ^{1/2}$ \\
period & $T=\dfrac{2\pi }{\xi ^{1/2}O_{1}}~\dfrac{R_{\circledS }}{c~\xi }$%
\end{tabular}
&
\begin{tabular}{ll}
size & $X=R_{\circledS }O_{1}~\varepsilon \sim 60\text{cm}$ \\
atom velocity & $v_{g}=\eta O_{1}\sim 20\text{cm}\text{s}^{-1}/c$ \\
Drift time & $2T_{D}=X/v_{g}=R_{\circledS }~\dfrac{\varepsilon }{\eta }\sim 3%
\text{s}$%
\end{tabular}
\\ \hline\hline
\end{tabular}
\\
Table 2.~:~definition of $O_{1},\xi ,\varepsilon $ and $\eta $%
\end{tabular}
\end{center}

With $\xi \simeq 0.9\ $one finds

\begin{center}
\begin{tabular}{c}
\begin{tabular}{|l|l|l|l|l|l|l|}
\hline
& $O_{1}$ & $\varepsilon $ & $\eta $ & $r$ & $v$ & $T$ \\ \hline
Earth & $2.6~10^{-5}$ & $3.6~10^{-3}$ & $2.7~10^{-5}$ & $7000\text{km}$ & $%
2.9~10^{-5}$ & $5900\text{s}$ \\ \hline
Jupiter & $1.4~10^{-4}$ & $6.0~10^{-5}$ & $4.8~10^{-6}$ & $78400\text{km}$ &
$1.5~10^{-5}$ & $12300\text{s}$ \\ \hline
\end{tabular}
\\
Table 3.%
\end{tabular}
\end{center}

\subsection{Comoving non rotating coordinates}

We consider the following tetrad, $e_{\hat{\sigma}}^{\alpha },$ comoving
with $O:$

\begin{eqnarray}
e_{\hat{0}}^{0} &=&u^{0}=1+\dfrac{\vec{v}^{2}}{2}+U\;+O_{4},\;\;e_{\hat{0}%
}^{k}=u^{k}=\left( 1+\dfrac{\vec{v}^{2}}{2}+U\right) v^{k}+\xi ^{2}O_{4}
\notag \\
e_{\hat{k}}^{0} &=&\left( 1+\dfrac{\vec{v}^{2}}{2}+U\right) \,v^{k}+\gamma
U\,v^{k}-g_{0k}+\xi ^{2}O_{4}  \label{tetrad} \\
e_{\hat{k}}^{j} &=&\delta _{k}^{j}+\dfrac{1}{2}v^{j}v^{k}+\dfrac{1}{2}\gamma
U\,\delta _{k}^{j}+\xi ^{2}O_{4}  \notag
\end{eqnarray}

The local metric is derived from \ref{local} with the change in the
notations $\left( \alpha \right) \rightarrow \hat{\alpha}$ and $u_{\left(
\sigma \right) }^{\alpha }\rightarrow e_{\hat{\sigma}}^{\alpha }.$

We limit the expansion of the metric at order $\varepsilon ^{2}\,\xi
^{3/2}O_{6};$ therefore we consider only the linear expression of the
Riemann tensor (eq.\,(\ref{riem})) and we assume that the free
fall is under control~: $\left\Vert \vec{a}\right\Vert <<\varepsilon
O_{3}\xi ^{3/2}\times \dfrac{c^{2}}{X}$ (\textit{i.e.} $\left\Vert \vec{a}%
\right\Vert <<8\text{m}\text{s}^{-2}$ for the Earth, $\left\Vert \vec{a}%
\right\Vert <<21\text{m}\text{s}^{-2}$ for Jupiter which is not very
restrictive) therefore we neglect the term $\left( \vec{a}\cdot \vec{X}%
\right) ^{2}$ in the metric (\ref{local}). One finds

\begin{eqnarray}
G_{\hat{0}\hat{0}} &=&1+2\,\vec{a}\cdot \vec{X}-\hat{U},_{\hat{k}\hat{\jmath}%
}\,X^{\hat{k}}X^{\hat{\jmath}}-\dfrac{1}{3}\hat{U},_{\hat{k}\hat{\jmath}\hat{%
\ell}}\,X^{\hat{k}}X^{\hat{\jmath}}X^{\hat{\ell}}+\varepsilon ^{2}\,\xi
^{3}O_{6}  \label{metun} \\
G_{\hat{0}\hat{m}} &=&-\left\{ \vec{\Omega}_{0}\wedge \vec{X}\right\} ^{\hat{%
m}}+\varepsilon ^{2}\xi ^{5/2}O_{5}\text{ and \ }G_{\hat{n}\hat{m}}=\eta _{%
\hat{n}\hat{m}}+\varepsilon ^{2}\xi ^{2}O_{4}  \label{metdeux}
\end{eqnarray}%
where $\vec{\Omega}_{0}$ is given below (see eq.\,(\ref{omzero})) while
the expressions such as $\hat{U},_{\hat{k}\hat{\jmath}}$ are nothing but $%
\left( U_{,mn}\,e_{\hat{k}}^{m}e_{\hat{\jmath}}^{n}\right) _{O}.$ The
position of the observer changes with time, therefore this quantity is a
function of $T.$

We did not consider the time dependence of the potential $U.$ One can prove
that it is correct when $\dfrac{\Delta U}{U}\times \dfrac{r}{cT_{c}}<<\xi
O_{2}$ where $\dfrac{\Delta U}{U}$ is the relative change of the potential
during the time $T_{c},$ at the distance $r$ of the origin.
This is generally the case.

In $G_{\hat{0}\hat{0}},$ the accuracy is limited to the terms of order of $%
\varepsilon ^{2}\,\xi ^{3}O_{6}.$ One can check that in such a case, the
approximation $e_{\hat{k}}^{m}=\delta _{\hat{k}}^{m}$ is valid therefore $%
\hat{U},_{\hat{k}\hat{\jmath}}\simeq \left( U_{,kj}\right) _{O}$ and $\hat{U}%
,_{\hat{k}\hat{\jmath}\hat{\ell}}=\left( U_{,kj\ell }\right) _{O}.$ The same
holds true for $\vec{\Omega}_{0}$ \textit{i.e.} $\left( \vec{\Omega}%
_{0}\right) ^{\hat{k}}\simeq \left( \vec{\Omega}_{0}\right) ^{k}$ (see
eq.\,(\ref{omzero})). Therefore, one can identify the space vectors $%
\overrightarrow{e}_{\hat{k}}$ of the tetrad and the space vectors $%
\overrightarrow{\partial }_{k}$ of the natural basis associated to the
geocentric coordinates. This would not be valid with an higher accuracy
where terms smaller than $\varepsilon ^{2}\,\xi ^{3}O_{6}$ are considered.

Calculating $\vec{\Omega}_{0}$ one finds the usual following
expression \cite{misner}

\begin{eqnarray}
\vec{\Omega}_{0} &=&\vec{\Omega}_{LT}+\vec{\Omega}_{dS}+\vec{\Omega}_{Th}
\label{omzero} \\
\left( \vec{\Omega}_{LT}\right) ^{\hat{k}} &\simeq &\left( \dfrac{\vec{J}}{%
r^{3}}-\dfrac{3}{r^{3}}\left( \vec{J}\cdot \vec{n}\right) \,\vec{n}\;-\dfrac{%
\alpha _{1}}{4}\vec{\nabla}U\wedge \vec{w}\right) ^{k}  \label{omlt} \\
\left( \vec{\Omega}_{dS}\right) ^{\hat{k}} &\simeq &\left( \left( 1+\gamma
\right) \vec{\nabla}U\wedge \vec{v}\right) ^{k}\ \text{and}\ \ \ \ \left(
\vec{\Omega}_{Th}\right) ^{\hat{k}}\simeq \left( \dfrac{1}{2}\vec{v}\wedge
\dfrac{d\vec{v}}{dt}\right) ^{k}  \label{omdsth}
\end{eqnarray}%
$\vec{n}$ is the direction of the satellite (fig.\,\ref{figdef}), 
$\vec{\Omega}_{LT}$ is the Lense-Thirring angular velocity, $\vec{\Omega}%
_{dS}$\ and\ $\vec{\Omega}_{Th}$ are the de Sitter and the Thomas terms%
\footnote{
The Thomas term reads $\vec{\Omega}_{Th}=\dfrac{1}{2}\vec{v}\wedge
\vec{A}$ where $\vec{A}$ is the \textquotedblright
acceleration\textquotedblright . From the relativistic point of view, it
would be better to define the Thomas term with the local physical
acceleration, $\vec{A}\simeq \dfrac{d\vec{v}}{dt}-\vec{\nabla}U$,
rather than the acceleration, $\dfrac{d\vec{v}}{dt}$, relatively to the
geocentric frame.}~:

\begin{center}
\begin{tabular}{c}
\begin{tabular}{|l|l|}
\hline
$\Omega _{LT}$ & $\sim \dfrac{J_{\circledS }}{M_{\circledS }^{2}}~\xi
^{2}O_{4}\times \dfrac{c\xi }{R_{\circledS }}$ \\
\hline
$\Omega _{dS}\sim \Omega _{Th}$ & $\sim \xi ^{3/2}O_{3}\times \dfrac{c\xi }{%
R_{\circledS }}$\\
\hline
\end{tabular}
\\
Table 4.%
\end{tabular}
\end{center}

With $\xi \sim 0.9,$ one finds

\begin{center}
\begin{tabular}{c}
\begin{tabular}{|l|l|l|l|}
\hline
& $J_{\circledS }/M_{\circledS }^{2}$ & $\Omega _{LT} $ & $\Omega
_{dS}\sim \Omega _{Th}$ \\
\hline
Earth & 750 & $\sim $10$^{-14}\text{rad}\text{s}^{-1}$ & $\sim $10$^{-12}%
\text{rad}\text{s}^{-1}$ \\
\hline
Jupiter & 855 & $\sim $10$^{-12}\text{rad}\text{s}^{-1}$ & $\sim $10$^{-11}%
\text{rad}\text{s}^{-1}$\\
\hline
\end{tabular}
\\Table 5.%
\end{tabular}
\end{center}

\subsection{Aberration and deflection of the light}

In the satellite, the experimental set-up is tied to a telescope which
points towards a \textquotedblright fixed\textquotedblright\ star (see
figure \ref{asu}). We assume that the star is far enough for the parallax to
be negligible. However the light rays suffer a gravitational deflection from
the central body and an aberration which depends on the position and the
velocity of the satellite. These effects result in an angular apparent
velocity which must be compared to the Lense-Thirring effect.

\begin{center}
\begin{figure}[ht]
\centering\includegraphics[width = .75\linewidth]{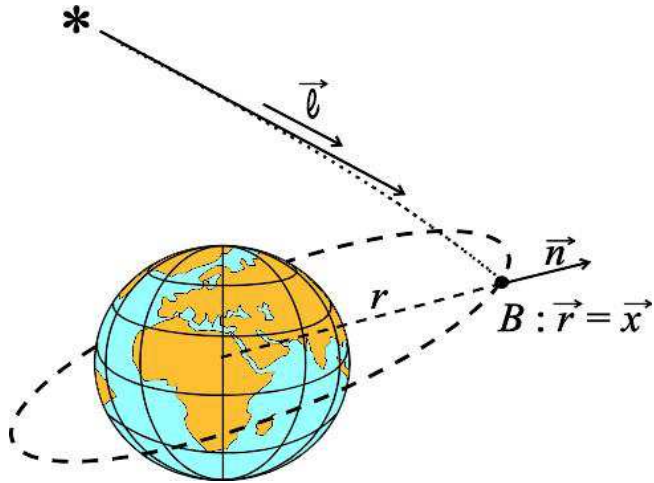}
\caption{\label{figdef} The deflection of the light.}
\end{figure}
\end{center}

In space time, the direction of the light from the star is given by the
4-vector $L_{\alpha }=\left\{ 1,\;\dfrac{\partial _{k}\,\varphi }{\partial
_{0}\,\varphi }\right\} $ where $\varphi $ is the phase of the light. In
order to calculate the phase $\varphi (t,x^{k})$ at point $\left\{
x^{k}\right\} $ and time $t$ we use the method which is summarized in
paragraph \ref{phasediff}. Now the line element is given by
eq.\,(\ref{ds2}), 
and $\omega $ is the angular frequency of the light at infinity.
These calculations are developed in another publication
\cite{Angonin}. Here, we just give the useful results.

The main gravitational contribution is due to the monopolar term of the
Newtonian potential~:%
\begin{equation}
L_{\alpha }=\left\{ 1,\;-\ell ^{k}+\left( 1+\gamma \right) \dfrac{M}{r}%
\dfrac{n^{k}-\ell ^{k}}{1-\vec{n}\cdot \vec{\ell}}+\delta \ell ^{k}+\delta
L^{k}\right\}  \label{lalfa}
\end{equation}%
where $\overrightarrow{\ell }$ is the unitary vector of figure \ref{figdef}
and $\overrightarrow{n}=\overrightarrow{r}/r$ .

$\bullet \qquad $The term $\delta \ell ^{k}$ is due to the quadrupolar term
of the central mass. This term is of order $J_{2}O_{2}$ when $%
\overrightarrow{\ell }$ is nearly orthogonal to the plane of the orbit.

$\bullet \qquad $The term due to $\dfrac{1}{2}\alpha _{1}U\,w^{k},$ a part
of $g_{0k}$ in the metric (\ref{ds2}), results in the modification $%
M_{\circledS }\rightarrow M=M_{\circledS }\left( 1-\dfrac{\alpha _{1}\,\vec{w%
}\cdot \vec{\ell}}{2\left( 1+\gamma \right) }\right) .$

$\bullet \qquad $The contribution due to the rotation of the central body is
of order of $J_{\circledS }/r^{2}$. The corresponding angular velocity is of
order of $J_{\circledS }/r^{2}/T\sim \dfrac{J_{\circledS }}{M_{\circledS
}^{2}}\dfrac{\xi ^{7/2}O_{5}}{2\pi }\dfrac{c}{R_{\circledS }}\sim \dfrac{\xi
^{1/2}O_{1}}{2\pi }\Omega _{LT}<<\Omega _{LT}$. It is negligible. The same
conclusion holds for the term $\Delta $ in eq.\,(\ref{4pole}).

$\bullet \qquad $The Sun, the satellites and the other planets, give a
contribution due to $U_{\ast }$ in (\ref{4pole}); it varies slowly with the
time and it is negligible, especially within the framework of a Fourier
analysis at a much higher frequency. An exception concerns the two
satellites of Jupiter, Andrastea and Metis whose period is approximately $%
25~10^{3}\text{s}$ which is the order of the period of a satellite on a low
orbit. However their mass do not exceed $10^{17}\text{kg}$ and the
gravitational deflections remain completely negligible.

For the observer $O,$ the space direction of the light is the four vector $%
\lambda ^{\alpha }=L^{\alpha }-L_{\beta }\,u^{\beta }\,u^{\alpha }.$ The
components of $\lambda ^{\alpha }$ relatively to the tetrad are $\left\{
\lambda ^{\hat{\alpha}}\right\} =\left( 0,\overrightarrow{\lambda }\right) .$
We define $\vec{\Lambda}=\Lambda \,\vec{\lambda}$ such as $-\Lambda _{\alpha
}\Lambda ^{\alpha }=\vec{\Lambda}\cdot \vec{\Lambda}=1.$

The tetrad (\ref{tetrad}) is especially useful to catch the orders of
magnitude of the various terms involved. However it is not the comoving
tetrad that we are looking for because the telescope that points towards the
far away star rotates relatively to this tetrad. The angular velocity of the
telescope relatively to $\left\{ e_{\hat{k}}^{\alpha }\right\} $ is $\vec{%
\Omega}_{\ast }=\vec{\Lambda}\wedge \dfrac{d\vec{\Lambda}}{dt}.$
Straightforward calculations give%
\begin{eqnarray}
\left( \vec{\Omega}_{\ast }\right) ^{\hat{k}} &=&-\left( \vec{\ell}\wedge
\dfrac{d\vec{v}}{dt}\right) ^{k}+\left( \vec{v}\wedge \dfrac{d\vec{v}}{dt}%
\right) ^{k}\notag\\
&&-\dfrac{3}{2}\left( \vec{\ell}\cdot \vec{v}\right) \left( \vec{%
\ell}\wedge \dfrac{d\vec{v}}{dt}\right) ^{k}+\dfrac{1}{2}\left( \vec{\ell}%
\cdot \dfrac{d\vec{v}}{dt}\right) \left( \vec{\ell}\wedge \vec{v}\right) ^{k}
\label{abrot} \\
&&-\dfrac{M}{r^{2}}\dfrac{1+\gamma }{1-\vec{n}\cdot \vec{\ell}}\left( \left(
\vec{\ell}\wedge \vec{v}\right) ^{k}+\left( \vec{\ell}\wedge \vec{n}\right)
^{k}\;\left[ \dfrac{\vec{\ell}\cdot \vec{v}\;-\vec{n}\cdot \vec{v}}{1-\vec{n}%
\cdot \vec{\ell}}-\vec{n}\cdot \vec{v}\right] \right)  \notag \\
&&+\left( \vec{\ell}\wedge \dfrac{d\delta \vec{\ell}}{dt}\right) ^{k}+\dfrac{%
1}{r}\times \xi ^{2}O_{4}  \notag
\end{eqnarray}%
Let us notice that we neglect the terms of order $\dfrac{1}{r}\times \xi
^{2}O_{4}$, which are much smaller than the Lense-Thirring angular velocity
because $\dfrac{J_{\circledS }}{M_{\circledS }^{2}}>>1$ and $\xi \sim 1.$

\subsection{Local coordinates tied to the telescope}

Now we introduce the tetrad tied to the telescope and the
interferometer $u_{\left( \sigma \right) }^{\alpha }$. It is
obtained from $e_{\hat{\rho}}^{\alpha }$ through a pure space rotation (%
\textit{i.e}. $u_{\left( 0\right) }^{\alpha }=e_{\hat{0}}^{\alpha
}=u^{\alpha })$ and whose vector $u_{\left( 1\right) }^{\alpha }$ points
towards the far away star ($u_{\left( 1\right) }^{\alpha }=-\Lambda ^{\alpha
})$.

At the required accuracy, it is possible to give a\ description of the Hyper
project with the Newtonian concept of space.

The rotation of the tetrad $\left\{ u_{\left( \sigma \right) }^{\alpha
}\right\} $ relatively to $\left\{ e_{\hat{\rho}}^{\alpha }\right\} $ is
characterized by the most general angular velocity $\vec{\Omega}_{u/e}=\vec{%
\Omega}_{\ast }-\varpi \vec{\Lambda}$ where $-\varpi \vec{\Lambda}$ is an
arbitrary angular velocity around the apparent direction of the star. The
change of the tetrad $u_{\left( \sigma \right) }^{\alpha
}\longleftrightarrow e_{\hat{\rho}}^{\alpha }$ is just an ordinary change of
basis in the space of the observer $O.$ In this transformation, $dT,$ $G_{%
\hat{0}\hat{0}}=G_{\left( 0\right) \left( 0\right) }$, $G_{\hat{0}\hat{m}%
}dX^{\hat{m}}=G_{\left( 0\right) \left( k\right) }dX^{\left( k\right) }$ and
$G_{\hat{n}\hat{m}}dX^{\hat{m}}dX^{\hat{n}}=G_{\left( j\right) \left(
k\right) }dX^{\left( j\right) }X^{\left( k\right) }$ behave as scalars. We
obtain the local metric from eqs.\,(\ref{metun}) and (\ref{metdeux}).
Then, using the expression \ref{psipsi} of $\Psi ,$ a straight forward
calculation gives~:%
\begin{eqnarray}
\Psi &=&2\,\vec{a}\cdot \vec{X}-\hat{U},_{\left( k\right) \left( j\right)
}\,X^{\left( k\right) }X^{\left( j\right) }-\dfrac{1}{3}\hat{U},_{\left(
k\right) \left( j\right) \left( \ell \right) }\,X^{\left( k\right)
}X^{\left( j\right) }X^{\left( \ell \right) } \\
\nonumber&&-2\underset{\left( k\right) }{\sum }\left\{ \left( \vec{\Omega}_{0}+\vec{%
\Omega}_{\ast }\right) \wedge \vec{X}\right\} ^{\left( k\right)
}v_{g}^{\;\left( k\right) }+\varepsilon ^{2}\xi ^{3}O_{6}
\end{eqnarray}%
with $\left\{ \left( \vec{\Omega}_{0}+\vec{\Omega}_{\ast }\right) \wedge
\vec{X}\right\} \cdot \overrightarrow{v}_{g}=\left\{ \left( \vec{\Omega}%
_{LT}-\left( \vec{\ell}\wedge \dfrac{d\vec{v}}{dt}\right) \right) \wedge
\vec{X}\right\} \cdot \overrightarrow{v}_{g}+\eta \varepsilon \xi
^{5/2}O_{5}.$

The Lense-Thirring contribution to $\Psi $ is $\Psi _{LT}\sim \dfrac{%
J_{\circledS }}{M_{\circledS }^{2}}\xi ^{3}\varepsilon \eta O_{6}.$
Therefore, within the present framework, the expected accuracy is of order of

\begin{center}
\begin{tabular}{c}
$
\begin{tabular}{|l|l|l|}
\hline
& Earth & Jupiter \\
\hline
$\dfrac{\varepsilon ^{2}\xi ^{3}O_{6}}{\left( J_{\circledS }/M_{\circledS
}^{2}\right) \xi ^{3}\varepsilon \eta O_{6}}\sim \dfrac{\varepsilon }{\left(
J_{\circledS }/M_{\circledS }^{2}\right) \eta }$ & 18\% & 1.5\%\\
\hline
\end{tabular}
$ \\
Table 6.%
\end{tabular}
\end{center}

\section{The phase shift}

Let us assume that any quantity can be known with an accuracy $\mu \sim
10^{-4}.$ This condition is not restrictive for the orbital parameters and
does not seem out of the present possibilities as far as the geometry of the
experimental device.

We consider that $\Psi $ is the amount of two terms, $\Psi _{k}$ and $\Psi
_{u}~:$ the term $\Psi _{k}$ is known; it can be modelled with the required
accuracy while $\Psi _{u}$ is unknown. The terms $\Psi _{k}$ fulfills the
condition $\mu \times \Psi _{k}\lesssim \varepsilon ^{2}\,\xi ^{3}O_{6}.$
With the previous orders of magnitude one finds

\begin{center}
\begin{tabular}{r}
\begin{tabular}{|c|c|c|c|}
\hline
&  & Earth & Jupiter \\ \hline
$\dfrac{\mu \hat{U},_{\left( k\right) \left( j\right) }\,X^{\left( k\right)
}X^{\left( j\right) }}{\varepsilon ^{2}\xi ^{3}O_{6}}\sim $ & $\dfrac{\mu }{%
\xi O_{2}}\sim $ & $1.6~10^{5}\in \Psi _{u}$ & $5.7~10^{3}\in \Psi _{u}$ \\
\hline
$\dfrac{\mu \hat{U},_{\left( k\right) \left( j\right) \left( \ell \right)
}\,X^{\left( k\right) }X^{\left( j\right) }X^{\left( \ell \right) }}{%
\varepsilon ^{2}\xi ^{3}O_{6}}\sim $ & $\dfrac{\mu \varepsilon }{\xi
^{1/2}O_{1}}\sim $ & $1.5~10^{-2}\in \Psi _{k}$ & $4.5~10^{-5}\in \Psi _{k}$
\\ \hline
$\dfrac{\mu \left\{ \left( \vec{\ell}\wedge \dfrac{d\vec{v}}{dt}\right)
\wedge \vec{X}\right\} \cdot \overrightarrow{v}_{g}}{\varepsilon ^{2}\xi
^{3}O_{6}}\sim $ & $\dfrac{\mu \eta }{\varepsilon \xi O_{2}}\sim $ & $%
1.2~10^{3}\in \Psi _{u}$ & $4.5~10^{2}\in \Psi _{u}$ \\ \hline
$\dfrac{\mu \eta \varepsilon \xi ^{5/2}O_{5}}{\varepsilon ^{2}\xi ^{3}O_{6}}%
\sim $ & $\dfrac{\mu \eta }{\varepsilon \xi ^{1/2}O_{1}}\sim $ & $%
3~10^{-2}\in \Psi _{k}$ & $6~10^{-2}\in \Psi _{k}$ \\ \hline
\end{tabular}
\\
\multicolumn{1}{c}{Table 7.}%
\end{tabular}
\end{center}

$\Psi _{u}$ reads

\begin{equation}
\Psi _{u}=2\,\vec{a}\cdot \vec{X}-\hat{U},_{\left( k\right) \left( j\right)
}\,X^{\left( k\right) }X^{\left( j\right) }-2\left\{ \vec{\Omega}\wedge \vec{%
X}\right\} \cdot \overrightarrow{v}_{g}+\varepsilon ^{2}\,\xi ^{3}O_{6}/\mu
\label{psifinal}
\end{equation}%
where the contribution $\hat{U},_{\left( k\right) \left( j\right)
}\,X^{\left( k\right) }X^{\left( j\right) }$ needs to be defined with an
accuracy better than $\varepsilon ^{2}\,\xi ^{3}O_{6}/\mu .$ This implies
that any known perturbation $\delta U$ can be included in $\Psi _{k}$ when $%
\dfrac{\delta U}{U}$ does not exceed the value given in table 8 below~:

\begin{center}
\begin{tabular}{c}
\begin{tabular}{|l|l|l|}
\hline
& Earth & Jupiter \\
\hline
$\dfrac{\delta U}{U}\lesssim \dfrac{\varepsilon \xi O_{2}}{\mu }$ & $%
2~10^{-8}$ & $10^{-8}$\\
\hline
\end{tabular}
\\
Table 8.%
\end{tabular}
\end{center}

The quadrupolar contribution is bigger than $10^{-3},$ it cannot be
considered in known term, however higher multipole can be included in $\Psi
_{k}$ if the accuracy $\mu $ is smaller than $10^{-6}$\ for the Earth
instead of $10^{-4}$ and $10^{-9}$ for Jupiter. The accuracy $\mu \sim
10^{-6}$ remains a very difficult challenge as far as the geometry of the
set-up is concerned (A. Landragin, private communication).

With the same order of magnitude for $\mu $, we obtain

\begin{equation}
\left( \vec{\Omega}\right) ^{\left( k\right) }=\left( \vec{\Omega}%
_{LT}-\varpi \overrightarrow{\Lambda }-\overrightarrow{\Lambda }\wedge
\dfrac{d\overrightarrow{v}}{dt}\right) ^{\left( k\right) }
\end{equation}%
where $\vec{\Omega}_{LT}$ is deduced from (\ref{omlt}) and $\dfrac{d%
\overrightarrow{v}}{dt}\simeq \overrightarrow{a}-\overrightarrow{\nabla }%
U\simeq \overrightarrow{a}+\dfrac{M_{\circledS }}{r^{2}}\overrightarrow{n}.$

The satellites, such as the Moon for the Earth, can bring a contribution to $%
\hat{U},_{\left( k\right) \left( j\right) }\,X^{\left( k\right) }X^{\left(
j\right) }$ at the required level of accuracy but with such a value of $\mu
, $ it could be included in $\Psi _{k}.$

Of course when modelling $\Psi _{k},$ one must be sure that every quantity
is known at the required accuracy. This must hold for any value of $\mu ,$
this is a necessary condition. Therfore the following relation must hold
true~: $\dfrac{\Delta U}{U}\lesssim \varepsilon \xi O_{2}$

\begin{center}
\begin{tabular}{c}
\begin{tabular}{|l|l|l|}
\hline
& Earth & Jupiter \\
\hline
$\dfrac{\Delta U}{U}\lesssim \varepsilon \xi O_{2}$ & $2~10^{-12}$ & $%
10^{-12}$\\
\hline
\end{tabular}
\\
Table 9.%
\end{tabular}
\end{center}

Such an accuracy is not achieved for Jupiter. For the Earth, considering the
difference between the various models (Godard Earth model 9 and 10) it
appears that the values of the high order multipoles are neither known nor
consistent at the required level ($10^{-12}).$ One can hope that the lack of
precision on the $J_{kn}$ coefficients \footnote{notations of the paper
by Ch. Marchal in Bulletin du museum d'histoire naturelle, 4\`{e}me s\'{e}%
rie, section C \textbf{18, }517 (1996)} is not important for $n\neq 0$
because the diurnal rotation modulates the frequency of the corresponding
contribution in $\hat{U},_{\left( k\right) \left( j\right) }\,X^{\left(
k\right) }X^{\left( j\right) }$. However it is necessary to increase our
knowledge of the axisymmetrical potential of the central body in order that $%
\xi ^{k}~\Delta J_{k0}\lesssim \varepsilon \xi O_{2}$ where $\Delta J_{k0}$
is the uncertainty on $J_{k0}=J_{k}.$ Such a relation holds true for $k=2.$
It could be presently achieved with low values of $\xi $ (on high orbits)
but the Lense-Thirring effect is proportional to $\xi ^{3}$ (see table 4
above) and it seems impossible to measure the Lense-Thirring effect for $\xi
<<1$ in not to far a future.

This question is crucial but Hyper might, itself, bring an answer to this
question by the means of the time analysis. Now we forget these problems
because in the simple case that we consider the quadratic quantities do not
bring any contribution to the signal.

\subsection{Significant terms in $\Psi _{u}$}

We consider that the motion of the satellite takes place in the $\left(
x,y\right) $-plane while the vector $\overrightarrow{\ell }$ lies in the $%
\left( x,z\right) $-plane. We assume that the eccentricity, $e,$ does not
exceed $\xi ^{1/2}O_{1}$.

We define

\begin{eqnarray}
\vec{J} &=&J_{x}\,\vec{e}_{\hat{1}}+J_{y}\,\vec{e}_{\hat{2}}+J_{z}\,\vec{e}_{%
\hat{3}}\;,\;\;\vec{n}=\cos \theta \,\vec{e}_{\hat{1}}+\sin \theta \,\vec{e}%
_{\hat{2}} \\
-\vec{\ell} &=&\cos \alpha \,\vec{e}_{\hat{1}}+\sin \alpha \,\vec{e}_{\hat{3}%
}\;,\;\;\vec{w}=w_{x}\,\vec{e}_{\hat{1}}+w_{y}\,\vec{e}_{\hat{2}}+w_{z}\,%
\vec{e}_{\hat{3}}
\end{eqnarray}%
$\vec{J},$ $\vec{\ell}$ and $\vec{w}$ are constant vectors. The angle $%
\theta $ and the distance $r$ depend on the time $T$.

\begin{center}
\begin{figure}[ht]
\centering\includegraphics[width = .5\linewidth]{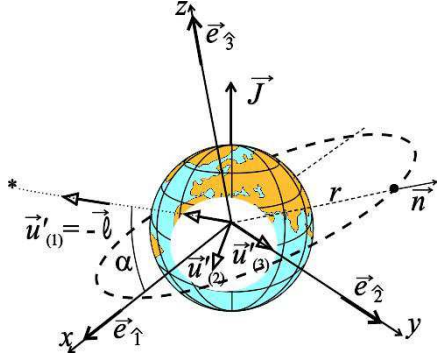}
\caption{\label{short3} The satellite and the fixed star}
\end{figure}
\end{center}

First we define the spatial triad $\vec{u}_{(n)}^{\prime }$~:
\begin{equation}
\vec{u}_{(1)}^{\prime }=\cos \alpha \,\overrightarrow{e}_{\hat{1}}+\sin
\alpha \overrightarrow{e}_{\hat{3}},\;\;\vec{u}_{(2)}^{\prime }=\sin \alpha
\,\overrightarrow{e}_{\hat{1}}-\cos \alpha \overrightarrow{e}_{\hat{3}},\;\;%
\vec{u}_{(3)}^{\prime }=\overrightarrow{e}_{\hat{2}}
\end{equation}

Let us outline that we have defined $\left( \vec{u}_{(1)}^{\prime }\right) ^{%
\hat{k}}=-\left( \vec{\Lambda}\right) ^{\hat{k}}+\xi ^{1/2}O_{1}$. Then, in
order to obtain the final tetrad $u_{\left( \sigma \right) }^{\alpha },$ we
perform an arbitrary rotation around $\vec{\Lambda}~:$
\begin{equation}
\begin{tabular}{l}
$\vec{u}_{(1)}=-\vec{\Lambda}=\vec{u}_{(1)}^{\prime }+\xi ^{1/2}O_{1}$ \\
$\vec{u}_{(2)}=\vec{u}_{(2)}^{\prime }\cos \sigma +\vec{u}_{(3)}^{\prime
}\sin \sigma +\xi ^{1/2}O_{1}$ \\
$\vec{u}_{(3)}=-\vec{u}_{(2)}^{\prime }\sin \sigma +\vec{u}_{(3)}^{\prime
}\cos \sigma +\xi ^{1/2}O_{1}$%
\end{tabular}
\label{uprimeau}
\end{equation}%
where $-\dfrac{d\sigma }{dT}=-\varpi $ is the angular velocity of the triad $%
\left\{ \vec{u}_{(k)}\right\} $ relatively to $\left\{ \vec{u}_{(k)}^{\prime
}\right\} .$

We can now assume that the experimental set-up is comoving with the triad $%
\vec{u}_{(n)}$ whose vector $\vec{u}_{(1)}$ points towards the fixed star.

During the flight of the atom, the quantity $\hat{U},_{\left( k\right)
\left( j\right) }$ in equation \ref{psifinal} does not remain constant
because the position of the satellite changes. One can consider that the
coordinate, $X=X^{\left( 1\right) }$ of the atom is a function of the time: $%
X=v_{g}\left( T-T_{0}\right) .$

Therefore we expand $\hat{U},_{\left( k\right) \left( j\right) }=\hat{U}%
,_{\left( k\right) \left( j\right) }(T_{0})+\hat{U},_{\left( k\right) \left(
j\right) \left( \ell \right) }\left( T_{0}\right) ~v^{\left( \ell \right)
}\times \dfrac{X}{v_{g}}$ where $v^{\left( \ell \right) }\overrightarrow{u}%
_{\left( \ell \right) }$ is the orbital velocity.

Before performing explicit calculation we notice that $\hat{U},_{\left(
k\right) \left( j\right) }\left( T_{0}\right) \,X^{\left( k\right)
}X^{\left( j\right) }$ will not bring any contribution to the phase
difference \ref{deltafi} because $O_{S}$ and $O_{S}^{\prime }$ are two
centers of symmetry. For $\mu <10^{-6}$ the term $\hat{U},_{\left( k\right)
\left( j\right) \left( \ell \right) }\left( T_{0}\right) ~\dfrac{v^{\left(
\ell \right) }}{v_{g}}\times X~X^{\left( k\right) }X^{\left( j\right) }\sim
\varepsilon ^{3}\xi ^{5/2}O_{5}/\eta $ can be included in $\Psi _{k}$ for
Jupiter ($\dfrac{\mu \varepsilon ^{3}\xi ^{5/2}O_{5}/\eta }{\varepsilon
^{2}\xi ^{3}O_{6}}<1)$, but it must considered for the Earth $\left( \dfrac{%
\mu \varepsilon ^{3}\xi ^{5/2}O_{5}/\eta }{\varepsilon ^{2}\xi ^{3}O_{6}}%
\sim 5>1\right) .$ However the quadrupole does not bring any contribution to
the phase difference that we calculate from $\Psi _{u}.$

In equation \ref{psifinal}, the spin $\varpi \vec{u}_{(1)}$ does not bring
any contribution in the term $\left\{ \vec{\Omega}\wedge \vec{X}\right\}
\cdot \overrightarrow{v}_{g}$ because $\vec{u}_{(1)},$ $\overrightarrow{X}$
and $\vec{v}_{g}$ are in the same plane.

Then, one obtains

\begin{equation}
\begin{tabular}{ll|l}
$\Psi _{u}=$ & $-2\left( \left( \dfrac{\vec{J}}{r^{3}}-\dfrac{3\left( \vec{J}%
\cdot \vec{n}\right) }{r^{3}}\vec{n}+\dfrac{\alpha _{1}\;M_{\circledS }}{%
4\,r^{2}}\,\vec{n}\wedge \vec{w}\right) \wedge \overrightarrow{X}\right)
\cdot \vec{v}_{g}$ & $A$ \\
& $-2\dfrac{M_{\circledS }}{r^{2}}\left( \left( \vec{u}%
_{(1)}\wedge \vec{n}\right) \wedge \overrightarrow{X}\right) \cdot \vec{v}%
_{g}+2\left( \left( \vec{u}_{(1)}\wedge \vec{a}\right) \wedge
\overrightarrow{X}\right) \cdot \vec{v}_{g}$ & $B$ \\
& $-\dfrac{6M_{\circledS }}{r^{4}}\left( \overrightarrow{X}\cdot \dfrac{%
\overrightarrow{v}}{v_{g}}\right) \left( \overrightarrow{X}\cdot
\overrightarrow{n}\right) X$ & $C$ \\
& $+2\,\vec{a}\cdot \overrightarrow{X}$ & $D$%
\end{tabular}
\label{psiu}
\end{equation}%
where $\overrightarrow{n}=\overrightarrow{n}\left( T_{0}\right) .$

In expression (eq.\,(\ref{psiu})) of $\Psi _{u}$ one can assume that $r=r_{0}$
is a constant because\ we assume that the excentricity is small $e\lesssim
O_{1},$ therefore the corrections are included in $\Psi _{k}.$

Moreover, for the same reason, one can drop the terms of order $O_{1}$ in
the expression of the tetrads. Therefore, it is clear that we can consider
the space as the ordinary space of Newtonian physics and that the usual
formulae to change the basis $\overrightarrow{\partial }_{k}$ into $%
\overrightarrow{e}_{\hat{k}}$ or $\overrightarrow{u}_{\left( k\right) }$are
valid.

In (\ref{psiu}), the terms of lines $A$ and $B$ are due to various
rotations~: respectively the Lense-Thirring rotation and the aberration. The
term of line $C$ is due to the displacement of the satellite during the
flight time of the atom and the term of line $D$ corresponds to some
residual acceleration due to the fact that point $O$ is not exactly in free
fall.

\subsection{The phase differences}

We use the expression (\ref{psiu}) of $\Psi _{u}$ in order to calculate $%
\delta \varphi $ given by \ref{deltafi}. We find

\begin{eqnarray}
\delta \varphi &=& -2\dfrac{mc}{\hbar \,r}S\,\left(
\dfrac{\vec{J}}{r_{0}^{2}}-\dfrac{3\left( \vec{J}\cdot \vec{n}\right) }{%
r_{0}^{2}}\vec{n}+\dfrac{\alpha _{1}\;M_{\circledS }}{4\,r_{0}}\,\vec{n}%
\wedge \vec{w}\right) \cdot \vec{u}_{(2)} \nonumber\\
&&  -2\dfrac{mc}{\hbar r}S\,\dfrac{M_{\circledS }}{r_{0}}%
\,\left( \vec{u}_{(1)}\wedge \vec{n}\right) \cdot \vec{u}_{(2)}-2\dfrac{mc}{%
\hbar }S\left( \vec{u}_{(1)}\wedge \dfrac{\overrightarrow{a}}{c^{2}}\right)
\cdot \vec{u}_{(2)} \\
&&  -\dfrac{4\pi \left( cT_{D}\right) ^{2}}{\lambda }%
\,\left( \vec{u}_{(3)}\cdot \dfrac{\vec{a}_{O_{S}}}{c^{2}}\right) \nonumber  \\
&&  -\dfrac{mc}{2\,\hbar r}S\,\dfrac{\left( cT_{D}\right)
^{2}}{r_{0}^{2}}\dfrac{M_{\circledS }}{r}\;\left( \left( \vec{%
u}_{(1)}\cdot \vec{n}\right) \left( \vec{u}_{(3)}\cdot \dfrac{%
\overrightarrow{v}}{c}\right) +\left( \vec{u}_{(3)}\cdot \vec{n}\right)
\left( \vec{u}_{(1)}\cdot \dfrac{\overrightarrow{v}}{c}\right) \right)
\nonumber %
\end{eqnarray}%
where $S=\dfrac{4\pi \hbar }{\lambda \,m}v_{g}T_{D}^{\;2}$ is the area of
the Sagnac loop. As we mentionned before, $\vec{J}$ and $M_{\circledS
}$ are expressed in geometrical units.

The two interferometers of the same ASU are assumed to lie in the same plane
but not necessarily with their center of symmetry $O_{S}$ and $O_{S}^{\prime
}$ at the same point. Therefore adding and subtracting the phase differences
delivered by the two interferometers one finds the two basic quantities
which are measured by the set-up \textit{i.e}.~: $%
{\mu}%
_{1}=\dfrac{1}{2}\left( \delta \varphi ^{\prime }-\delta \varphi \right) $
and $%
{\mu}%
_{2}=\dfrac{1}{2}\left( \delta \varphi ^{\prime }+\delta \varphi \right) .$

We define the shift $\overrightarrow{\delta }=\overrightarrow{X}%
_{O_{S}^{\prime }}-\overrightarrow{X}_{O_{S}}$ and the acceleration $%
\overrightarrow{a}=\dfrac{1}{2}\left( \vec{a}_{O_{S}}+\,\vec{a}%
_{O_{S}^{\prime }}\right) $ where $\vec{a}_{O_{S}}$ and $\,\vec{a}%
_{O_{S}^{\prime }}$ are the accelerations at point $O_{S}$ and $%
O_{S}^{\prime }.$ We drop several terms which can be included into $\Psi
_{k}.$ Then we obtain the quantities which can be measured~:
\begin{eqnarray}
{\mu}
_{1}+\dfrac{2v_{g}}{c}
{\mu}
_{2} &=&\dfrac{8\pi }{\lambda }\left( cT_{D}\right) ^{2}\left\{ \dfrac{%
M_{\circledS }}{r_{0}^{2}}\vec{u}_{(3)}\cdot \vec{n}+\vec{\Omega}_{LT}\cdot
\vec{u}_{(2)}\right\} \dfrac{v_{g}}{c} \\
&&-\dfrac{2\pi \left( cT_{D}\right) ^{2}}{\lambda r_{0}}\dfrac{M_{\circledS }%
}{r_{0}}\left( \vec{u}_{(3)}\cdot \dfrac{\overrightarrow{\delta }}{r}%
-3\left( \vec{n}\cdot \vec{u}_{(3)}\right) \left( \vec{n}\cdot \dfrac{%
\overrightarrow{\delta }}{r}\right) \right) \nonumber\\
&&-\dfrac{2\pi }{\lambda }\left( cT_{D}\right) ^{2}\left( \dfrac{v_{g}T_{D}}{%
r_{0}}\right) ^{2}\dfrac{M_{\circledS }}{r_{0}^{2}}\left\{ \left( \vec{u}%
_{(1)}\cdot \vec{n}\right) \left( \vec{u}_{(3)}\cdot \dfrac{\overrightarrow{v%
}}{c}\right) \right. \nonumber\\
&&\left. +\left( \vec{u}_{(3)}\cdot \vec{n}\right) \left( \vec{u}_{(1)}\cdot
\dfrac{\overrightarrow{v}}{c}\right) \right\} \nonumber\\
{\mu}%
_{2} &=&\dfrac{1}{2}\left( \delta \varphi ^{\prime }+\delta \varphi \right)
=-\dfrac{4\pi \left( cT_{D}\right) ^{2}}{\lambda }\left\{ \dfrac{%
\overrightarrow{a}}{c^{2}}\cdot \vec{u}_{(3)}\right\}
\end{eqnarray}

\subsection{Discussion}

We define $\alpha$ as the direction of the fixed star
(fig.\,(\ref{short3})),
 and the projection, $\overrightarrow{J}_{\shortparallel }$ of $%
\overrightarrow{J}$ on the plane of the orbit~:

$\overrightarrow{J}_{\shortparallel }=J_{\shortparallel }\left( \cos \theta
_{J}\;\overrightarrow{e}_{\hat{1}}+\sin \theta _{J}\;\overrightarrow{e}_{%
\hat{2}}\right) $ and $\overrightarrow{w}_{\shortparallel
}=w_{\shortparallel }\left( \cos \theta _{w}\;\overrightarrow{e}_{\hat{1}%
}+\sin \theta _{w}\;\overrightarrow{e}_{\hat{2}}\right) .$ Then%
\begin{eqnarray}
{\mu}%
_{1}+2\dfrac{v_{g}}{c}%
{\mu}%
_{2} &=&\dfrac{2\pi \left( cT_{D}\right) ^{2}}{\lambda r_{0}}\times \left\{
K_{0}+K_{\sigma }+K_{2\sigma }+K_{2\theta }\right. \\
&&\left. +K_{\theta -\sigma }+K_{\theta +\sigma }+K_{2\theta -\sigma
}+K_{2\theta +\sigma }+K_{2\theta -2\sigma }+K_{2\theta +2\sigma
}\right\} \nonumber
\end{eqnarray}%
with%
\begin{eqnarray}
K_{0} &=&\dfrac{M_{\circledS }}{4\,r_{0}}\left( 3\sin ^{2}\alpha -1\right)
\times \dfrac{\delta ^{(3)}}{r_{0}} \\
K_{\sigma } &=&\dfrac{v_{g}}{c}\left\{ \left[ \left( 1-\sin \alpha \right)
\cos \left( \sigma +\theta _{J}\right) -\left( 1+\sin \alpha \right) \cos
\left( \sigma -\theta _{J}\right) \right] \times \dfrac{J_{\shortparallel }}{%
r_{0}^{\;2}}\right. \nonumber\\
&&\left. -4\cos \alpha \,\cos \sigma \times \dfrac{J^{\hat{3}}}{r_{0}^{\;2}}%
\right\} -\dfrac{3M_{\circledS }}{2r_{0}}\cos \alpha \sin \alpha \sin \sigma
\times \dfrac{\delta ^{\left( 1\right) }}{r_{0}} \\
K_{2\sigma } &=&\dfrac{3M_{\circledS }}{4r_{0}}\left( 1-\sin ^{2}\alpha
\right) \,\left[ \sin \left( 2\sigma \right) \times \dfrac{\delta ^{(2)}}{%
r_{0}}+\cos \left( 2\sigma \right) \times \dfrac{\delta ^{(3)}}{r_{0}}\right]
\\
K_{2\theta } &=&-\dfrac{3M_{\circledS }}{4r_{0}}\left( 1-\sin ^{2}\alpha
\right) \,\cos \left( 2\theta \right) \times \dfrac{\delta ^{\left( 3\right)
}}{r_{0}}
\end{eqnarray}%
\begin{eqnarray}
K_{\theta -\sigma } &=&\dfrac{M_{\circledS }}{r_{0}}\dfrac{v_{g}}{c}\times
\left\{ -2\left( 1+\sin \alpha \right) \,\sin \left( \theta -\sigma \right)
\right. \\
&&\left. +\dfrac{\alpha _{1}}{2}\cos \alpha \,\sin \left( \theta -\sigma
-\theta _{w}\right) \times \dfrac{w_{\shortparallel }}{c}\right. \nonumber\\
&&\left. +\dfrac{\alpha _{1}}{2}\left( 1+\sin \alpha \right) \sin \left(
\theta -\sigma \right) \times \dfrac{w^{\hat{3}}}{c}\right\} \nonumber\\
K_{\theta +\sigma } &=&\dfrac{M_{\circledS }}{r_{0}}\dfrac{v_{g}}{c}\times
\left\{ -2\left( 1-\sin \alpha \right) \,\sin \left( \theta +\sigma \right)
\right. \\
&&\left. +\dfrac{\alpha _{1}}{2}\cos \alpha \,\sin \left( \theta +\sigma
-\theta _{w}\right) \times \dfrac{w_{\shortparallel }}{c}\right. \nonumber\\
&&\left. -\dfrac{\alpha _{1}}{2}\left( 1-\sin \alpha \right) \sin \left(
\theta +\sigma \right) \times \dfrac{w^{\hat{3}}}{c}\right\} \nonumber
\end{eqnarray}%
\begin{eqnarray}
K_{2\theta -\sigma } &=&-3\dfrac{v_{g}}{c}\left( 1+\sin \alpha \right) \cos
\left( 2\theta -\sigma -\theta _{J}\right) \times \dfrac{J_{\shortparallel }%
}{r_{0}^{\;2}} \\
&&+\dfrac{3M_{\circledS }}{4r_{0}}\cos \alpha \,\left( 1+\sin \alpha \right)
\sin \left( 2\theta -\sigma \right) \times \dfrac{\delta ^{\left( 1\right) }%
}{r_{0}} \nonumber\\
&&-\left( \dfrac{M_{\circledS }}{r_{0}}\right) ^{3/2}\dfrac{cv_{g}T_{D}^{\;2}%
}{2\,r_{0}^{\;2}}\left( 1+\sin \alpha \right) \cos \left( 2\theta -\sigma
\right) \nonumber\\
K_{2\theta +\sigma } &=&3\dfrac{v_{g}}{c}\left( 1-\sin \alpha \right) \cos
\left( 2\theta +\sigma -\theta _{J}\right) \times \dfrac{J_{\shortparallel }%
}{r_{0}^{\;2}} \\
&&+\dfrac{3M_{\circledS }}{4r_{0}}\cos \alpha \,\left( 1-\sin \alpha \right)
\sin \left( 2\theta +\sigma \right) \times \dfrac{\delta ^{\left( 1\right) }%
}{r_{0}} \nonumber\\
&&-\left( \dfrac{M_{\circledS }}{r_{0}}\right) ^{3/2}\dfrac{cv_{g}T_{D}^{\;2}%
}{2\,r_{0}^{\;2}}\left( 1-\sin \alpha \right) \cos \left( 2\theta +\sigma
\right)\nonumber
\end{eqnarray}%
\begin{eqnarray}
K_{2\theta -2\sigma } &=&\dfrac{3M_{\circledS }}{8\,r_{0}}\left( 1+\sin
\alpha \right) ^{2} \\ &&\times\left\{ \sin \left( 2\theta -2\sigma \right) \times
\dfrac{\delta ^{\left( 2\right) }}{r_{0}}-\cos \left( 2\theta -2\sigma
\right) \times \dfrac{\delta ^{\left( 3\right) }}{r_{0}}\right\} \nonumber\\
K_{2\theta +2\sigma } &=&-\dfrac{3M_{\circledS }}{8\,r_{0}}\left( 1-\sin
\alpha \right) ^{2} \\ && \times \left\{ \sin \left( 2\theta +2\sigma \right) \times
\dfrac{\delta ^{\left( 2\right) }}{r_{0}}+\cos \left( 2\theta +2\sigma
\right) \times \dfrac{\delta ^{\left( 3\right) }}{r_{0}}\right\} \nonumber
\end{eqnarray}

Each of these terms, except $K_{0},$ has a specific frequency. They can be
measured and distinguished from each other.

The Lense-Thirring effect due to the angular momentum of the central body
appears in the terms $K_{\sigma }$ and $K_{2\theta \pm \sigma }$ while the
possible existence of a preferred frame appears in $K_{\theta \pm \sigma }$
which depends on the components of $\alpha _{1}\overrightarrow{w}.$

The signal due to the Lense-Thirring effect is associated with the signal
due to $\delta ^{\left( 1\right) }.$ Today, it seems impossible to reduce $%
\delta ^{\left( 1\right) }$ significantly, this is the reason why it should
be calculated from the Fourier analysis of the signal itself altogether with
the velocity $\alpha _{1}\overrightarrow{w}.$

The interest of the spin is obvious. If $\sigma$ is constant 
(no spin) the signal
is the sum of two periodic signals with frequency $\nu _{O}$ and $2\nu _{O}$
where $\nu _{O}$ is the orbital frequency of the satellite~; therefore one
ASU gives two informations (two functions of the time). When the satellite
spins, we get 9 functions of the time $t.$ The information is much more
important in this case.

\section{Conclusion}

In Hyper, the Lense-Thirring effect is associated with many perturbations
which cannot be cancelled. We have exhibited the various terms that one
needs to calculate in order to obtain the full signal and we have emphasized
the necessity to increase our knowledge of the Newtonian gravitational
potential. This is still more crucial for Jupiter despite the fact that the
Lense-Thirring effect is much bigger.

Using four parameters, $\xi ,$ $\varepsilon ,$ $\eta $ and $\mu $ defined in
table 2, we have also sketched a method to take into account the residual
gravitational field in a nearly free falling satellite, namely the tidal and
higher order effects.

Compared with GPB, the principle of the measure is not the same, the
difficulties are quite different but the job is not easier. For instance,
considering the quantities $K_{\sigma }$ or $K_{2\theta \pm \sigma }$ above,
one can check that, for an Earth satellite, $\delta ^{\left( 1\right) }$
must remain smaller than $2\text{nm}$\ for the corresponding signal to
remain smaller than the Lense-Thirring one. It does not seem that such a
precision can be controlled in the construction of the experimental device
itself. It is therefore necessary to measure $\delta ^{\left( 1\right) }$
with such an accuracy.

What can be deduced from the time analysis depends on the accuracy of the
various parameters. From $K_{2\theta \pm 2\sigma }$ we deduce $\alpha $.
Then from $K_{\theta \pm \sigma }$ we obtain $\alpha _{1}w_{\shortparallel
}/c$ and $\alpha _{1}w^{\hat{3}}/c$ as two functions of $\theta _{w}.$
Therefore one can check if $\alpha _{1}w=0$ or not.

From $K_{2\theta \pm \sigma }$ one can calculate $\dfrac{M_{\circledS }c}{%
r_{0}^{2}v_{g}}\delta ^{\left( 1\right) }$ as a function $G_{\left( +\right)
}$ of $J_{\shortparallel }/r_{0}^{2},$ $\theta _{J}$ and $\left( \dfrac{%
M_{\circledS }}{r_{0}}\right) ^{3/2}\dfrac{c^{2}T_{D}^{2}}{r_{0}^{2}}$\ and
as a different function, $G_{\left( -\right) }$ of the same arguments. One
could check the equality $G_{\left( +\right) }=G_{\left( -\right) }.$ If we
assume that $\overrightarrow{J}_{\circledS }$ is known, then $\theta _{J}$
is known and from the equality $G_{\left( +\right) }=G_{\left( -\right) }$
we deduce the value of $J_{\shortparallel }.$ Using the relation $%
J_{\shortparallel }=\dfrac{1+\gamma +\alpha _{1}/4}{2}\left( \overrightarrow{%
J}_{\circledS }\right) _{\shortparallel }$ one could check whether $\gamma
+\alpha _{1}/4=1.$

$K_{\sigma }$ would give $\dfrac{M_{\circledS }c}{r_{0}^{2}v_{g}}\delta
^{\left( 1\right) }$ as a function of $J_{\shortparallel }/r_{0}^{2},$ $%
\theta _{J}$ and $J^{\hat{3}}/r_{0}^{2}.$ Using the previous results, we
obtain $J^{\hat{3}}$. The relation $J^{\hat{3}}=\dfrac{1+\gamma +\alpha
_{1}/4}{2}\left( \overrightarrow{J}_{\circledS }\right) ^{\hat{3}}$ gives an
other test of the value of $\gamma +\alpha _{1}/4.$

But over all, the best test would be that the signal (as a function of the
time) fits the theoretical prediction.

As a final conclusion let us put forwards that the geometric scheme which
has been used is just a preliminary contribution to the discussion on the
feasibility of Hyper. Only a more powerful model can answer the question.
This model should take into account all the gravitational perturbations that
we have outlined here and it should consider the interaction between laser
fields and matter waves in more a realistic manner. Such an approach has
been recently developed \cite{Antoine}, \cite{Borde2004}
it could give definitive results in the future.

\bibliography{hyper}
\end{document}